\documentclass[a4paper,11pt]{article}

\usepackage{bm}
\usepackage{epsfig}
\usepackage{graphicx}
\usepackage{graphics}
\usepackage{color}

\usepackage{geometry}
\usepackage{comment}

\usepackage{amsmath}
\usepackage{amsfonts}
\usepackage{amssymb}

\usepackage{hyperref}

\begin{document}


\title{Existence of Anisotropic Spin Fluctuations at Low Temperature in the Normal Phase of the Superconducting Ferromagnet UCoGe}
\author{M. Taupin\footnote{SPSMS-INAC, UMR-E CEA-Universit\'{e} de Grenoble, 17 rue des Martyrs, 38054 Grenoble, France}, L. Howald\footnote{Physik-Institut der Universit\"{a}t Z\"{u}rich, Winterthurerstrasse 190, CH-8057 Z\"{u}rich, Switzerland}, D. Aoki$^{\ast,}$\footnote{Tohoku univ, Inst. Mat. Res., Oarai, Ibaraki 3111313, Japan}, J. Flouquet$^{\ast}$ and J.P. Brison$^{\ast}$}

\maketitle

E-mail: jean-pascal.brison@cea.fr

\begin{abstract}
Thermal conductivity measurements have been performed on the superconducting ferromagnet UCoGe down to very low temperature and under magnetic field. In addition to the electronic quasiparticle thermal conductivity, additional contributions to the thermal transport are detected: they are sensitive to the amplitude and direction of the magnetic field, and at low temperature, they display a strong anisotropy with the heat current direction. We identify these contributions as arising from magnetic fluctuations. Detection of such fluctuations on the thermal transport in 3D weak ferromagnets is very rare if not unique, and pledges for a strongly itinerant character of the magnetism of UCoGe.
\end{abstract}


The first superconducting ferromagnet, UGe$_2$, has been discovered more than ten years ago \cite{Saxena2000}. The field has been continuously expanding with the discovery of the zero pressure ferromagnetic superconductors URhGe \cite{Aoki2001} and UCoGe \cite{Huy2007}, as well as with the wealth of unconventional phenomena discovered when exploring their properties under pressure and magnetic fields \cite{Aoki2012a}. 
In UCoGe, the $f$-electrons are strongly involved in both magnetic and superconducting orders \cite{Ohta2008,Visser2009}, and the magnetic ground state on which superconductivity develops is a cornerstone for the supposed equal-spin pairing superconducting ground state, as well as for the field dependence of the superconducting properties \cite{Mineev2010}.

The difficulty to have a precise description of the ground state is usual in heavy-fermion systems, as rich physics does not arise in clear-cut limit cases, but in intermediate regimes where Kondo effect, itineracy of the \textit{f}-bands and intersite magnetic interactions compete on equal footings. Another difficulty comes specifically for these ferromagnetic superconductors, from the lack of microscopic data on the magnetic excitation spectrum at low temperature. UGe$_2$ is the best known system, with inelastic neutron data at zero pressure showing that it is intermediate between a localized and itinerant system \cite{Huxley2003,Raymond2004}. Under pressure, when the Curie temperature ($T_{Curie}$) is lowered together with the ordered moment, and superconductivity appears, it is likely that UGe$_2$ has gone closer to the itinerant electron limit case. 
For UCoGe, NMR has succeeded to demonstrate the strong Ising anisotropy of the static and dynamical properties \cite{Ihara2010}. It also revealed the presence of longitudinal magnetic fluctuations up to very high temperature (80K) compared to $T_{Curie}$ (around 2.5-2.8K) \cite{Ihara2010}, as well as a strong suppression of these fluctuations by a magnetic field applied along the easy magnetization axis \cite{Hattori2012}.

This paper is focused on the normal state properties of UCoGe, with the first study of thermal conductivity at low temperature and under magnetic field. We show that, unexpectedly, thermal transport reveals a contribution of magnetic excitations at high temperature, and below $T_{Curie}$, an anisotropic dispersion of these excitations. The results are in agreement with the itinerant character of the magnetism of UCoGe, even though it seems to be a rare case of weak ferromagnet where magnetic fluctuations act not only as a scattering mechanism, but also emerge as a proper channel for heat conduction. This feature may be due to the duality between the itinerant and localized character of the electrons.


UCoGe crystallizes in an orthorhombic structure. The \textbf{c}-axis is the easy magnetization axis, the \textbf{a}- \textbf{b}-axis being the hard and intermediate axis respectively. High-quality single crystals were grown by the Czhochralski method in a tetra-arc furnace, and further annealed. Thermal conductivity has been measured on 4 different bar-shaped samples, cut along the different crystallographic directions (2 along the \textbf{c}-axis, 1 along \textbf{a}, and 1 along \textbf{b}), with very different residual resistivity ratios (RRR, ranging from 16 up to 150), which allowed to distinguish pure electronic quasiparticle contributions, from others (phonons, spin fluctuations$\ldots$). They have been characterized by specific heat, Laue X-ray diffraction and resistivity measurements. The samples are labelled $S^{i}_{j}$, where the superscript \textit{i} indicates the current direction and the subscript  \textit{j} is the RRR. In the following, we call electronic ``quasiparticle contribution" to the thermal conductivity ($\kappa_{qp}$), that which is related to charge transport and appears in the Wiedemann-Franz law (WFL).
 
\begin{figure}[!ht]
	\centering
		\includegraphics[width=1\textwidth]{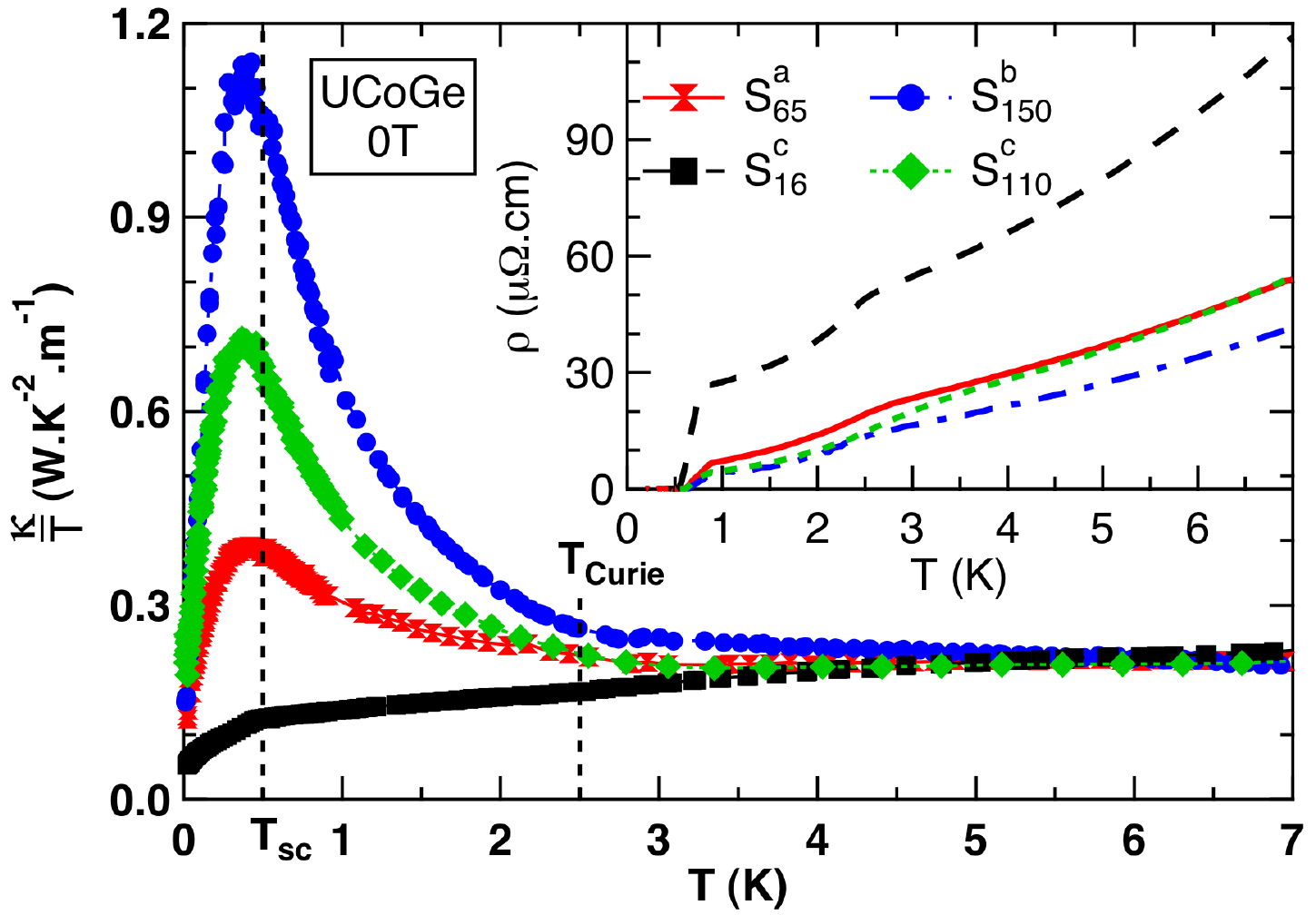}
	\caption{$\kappa/T$ at zero field of four crystals of UCoGe, with the heat current along the \textbf{a}, \textbf{b} and \textbf{c}-axis, with different RRR (see text for labelling). Note that $\kappa/T$ does increase with the RRR at 0.5K, but is almost independent of the RRR above 3K. \textit{Inset}: Resistivity. 
	}
	\label{fig:K_rho_0T}
\end{figure}
Low temperature thermal conductivity measurements have been performed down to 10mK and up to 8.5T with the standard one heater two thermometers setup. Moreover, results under a magnetic field applied along both the \textbf{c}-axis and the \textbf{b}-axis will be presented for samples $S^{c}_{16}$ and $S^{b}_{150}$. The electrical resistivity was measured simultaneously for all samples, so that the WFL could be checked for all samples ($L=\frac{\kappa\rho}{T}\to L_0=2.44.10^{-8}W.\Omega.K^{-2}$ when $T\to0K$, $L_0$ the Lorentz number and $\rho$ the resistivity). 
Results are independent from the applied temperature gradient, which was typically $\Delta T/T\sim 1-2\%$, except for sample $S^{b}_{150}$, where it was between $0.1$ and $1\%$ due to its higher conductivity.
As physical properties have a strong angular dependence \cite{Aoki2009}, the alignment of the samples in the magnetic field along the \textbf{b}-axis was adjusted in situ with two goniometers with piezo-electric actuators.

Figure \ref{fig:K_rho_0T} shows the thermal conductivity divided by the temperature ($\kappa/T$) at zero field of the four samples up to 7K. The ferromagnetic and superconducting transitions are shown by the vertical dashed lines. $\kappa/T$ shows a kink at $T_{Curie}$, and at the superconducting transition ($T_{sc}\approx0.5K$), displays a kink in sample $S^{c}_{16}$ and a much weaker feature on the other samples ($T_{sc}$ is slightly above the maximum of $\kappa/T$). For each crystal, $\kappa/T$ in the paramagnetic state (above 3K) varies very little whereas it increases strongly in the ferromagnetic state for samples with large RRR. By contrast, the resistivity (in inset) of all samples decreases monotonically in the same temperature range. So, above 3K, improvement of the electronic mean free path seems to have almost no effect on $\kappa/T$ (same value at 7K for all samples despite a factor three between the resistivity of samples $S^{c}_{16}$ and $S^{b}_{150}$ at this temperature). This is even true in the whole temperature range investigated for sample $S^{c}_{16}$, as $\kappa/T$ continuously decreases on cooling despite the decrease of the resistivity (suppression of inelastic scattering). 

\begin{figure}[!ht]
	\centering
		\includegraphics[width=1\textwidth]{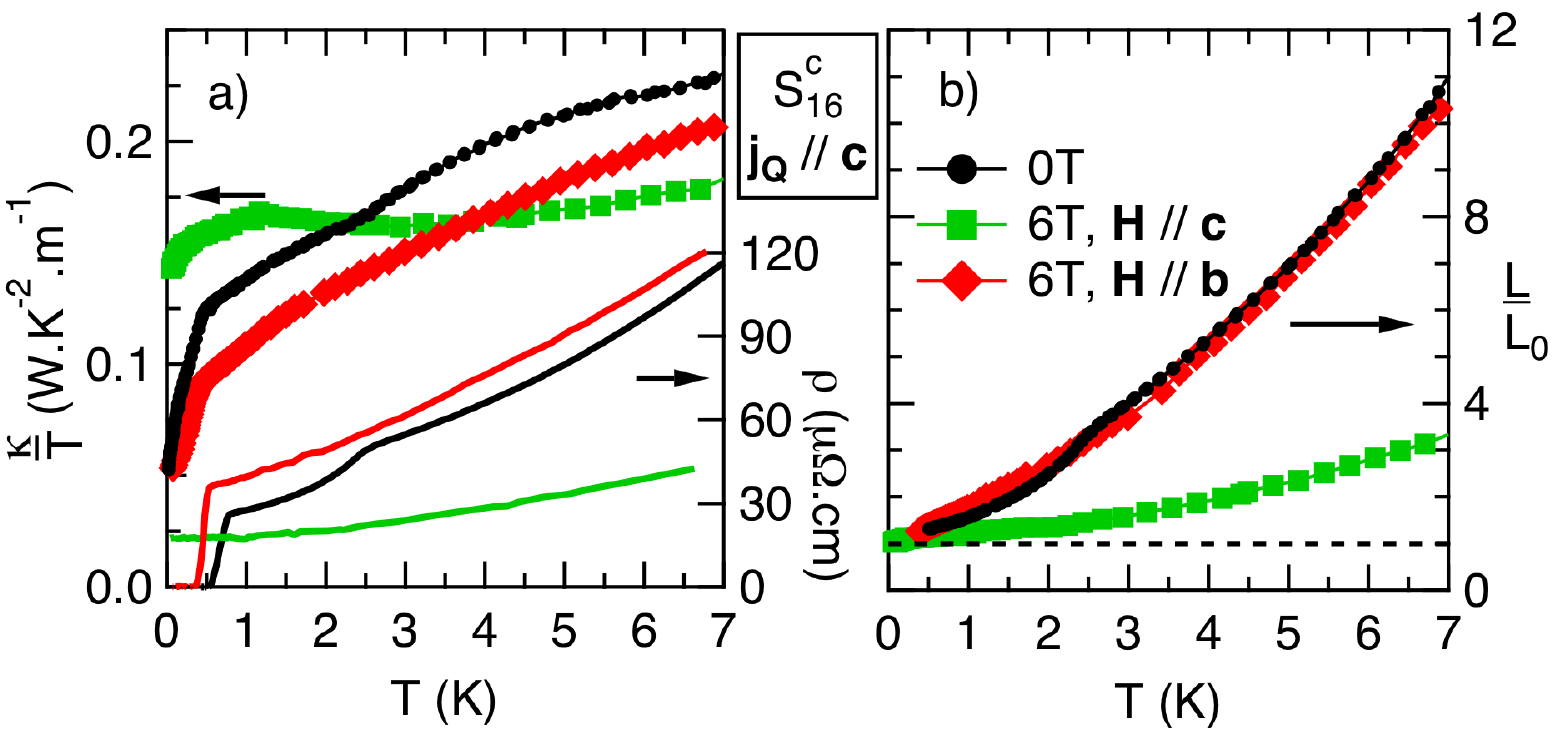}
	\caption{\textit{a)} Thermal conductivity and resistivity of sample $S^{c}_{16}$ at 0T and 6T for the field in both directions. \textit{b)} The corresponding Lorentz ratio, reaching 10 at 7K, but strongly suppressed for \textbf{H}=6T//\textbf{c}. Both point to an extra contribution to $\kappa/T$ above 3K, suppressed by a field if it is directed along the \textbf{c}-axis.} 
	\label{fig:kappa_A2}
\end{figure}
These opposite behaviors of charge and heat transport point to the existence of a sizable contribution to $\kappa/T$ in addition to the electronic quasiparticle contribution ($\kappa_{qp}$), dominant for all samples above 3K, and in the whole temperature range for sample $S^{c}_{16}$. We will see below that the usual lattice contribution ($\kappa_{ph}$) can only be part of this extra contribution: $\kappa_{ph}\approx bT^2$. The coefficient $b$ (temperature and field independent) was consistently estimated to be in the range $b\approx 0.01-0.02 W.K^{-3}.m^{-1}$ for the different samples, notably from the high field measurements, similar to estimates in other heavy fermion systems (see CeRhIn$_5$ for example \cite{Paglione2005}). We call extra contribution ($\kappa_{extra}$) the thermal conductivity after removal of the quasiparticle and phonon contributions, assuming that the contributions of the different channels are additive: $\kappa = \kappa_{qp} +\kappa_{ph}+\kappa_{extra}$. 

$\kappa_{extra}$ is best seen on the Lorentz ratio: $\frac{L}{L_{0}}=\frac{1}{L_{0}}.\frac{\kappa\rho}{T}$, displayed on Fig.\ref{fig:kappa_A2}b) for sample $S^{c}_{16}$ (\textbf{j}$_Q$//\textbf{c}): it reaches 1 when extrapolated at zero temperature, a good check of the validity of the measurements, but it is always superior to 1 at finite temperature, which confirms the presence of extra contributions, particularly dominant at high temperature ($\frac{L}{L_{0}}$ larger than 10 at 7K in zero field). The effect of the magnetic field is also visible on the same figure: for a field of 6T for \textbf{H}//\textbf{b}, there is almost no effect on the Lorentz ratio, whereas the same field applied along the \textbf{c}-axis reduces drastically $\frac{L}{L_{0}}$. 
\begin{figure}[!ht]
	\centering
		\includegraphics[width=1\textwidth]{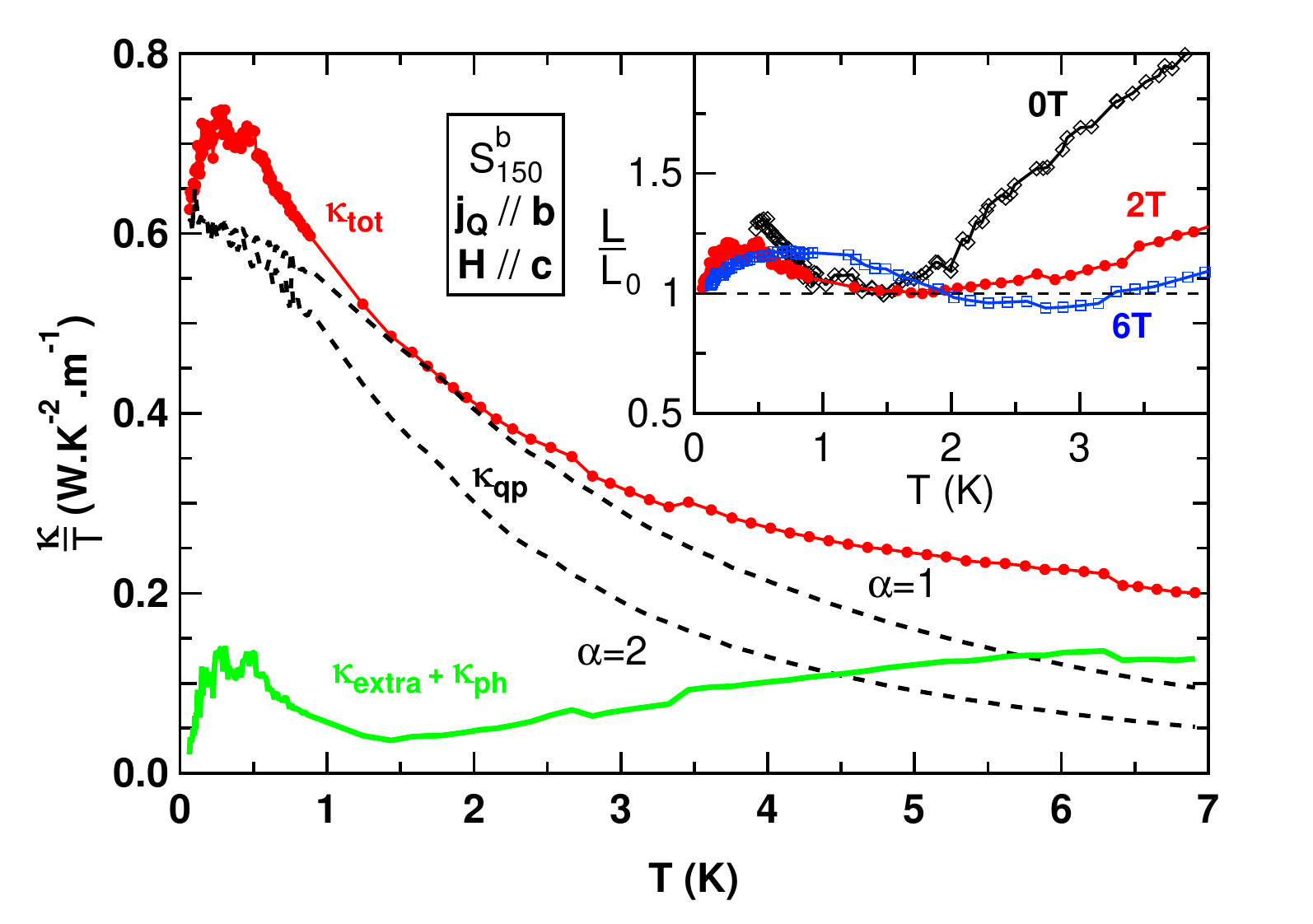}
	\caption{Total thermal conductivity (points) at 2T \textbf{H}//\textbf{c}, and electronic quasiparticle thermal conductivity (dashed lines) deduced via the extended WFL (with $\alpha=1$ or $2$, see text) of sample $S^{b}_{150}$. The lower curve (full line) shows the additional contribution ($\kappa_{extra}+\kappa_{ph}$), with a large almost model independent contribution below 0.5K. \textit{Inset}: the Lorentz ratio at 0T, 2T and 6T, \textbf{H}//\textbf{c}.}
	\label{fig:kappa_A12}
\end{figure}
A field of $6T$ along the \textbf{c}-axis (same Fig.\ref{fig:kappa_A2}a)) improves heat transport below $2K$, but reduces it above, whereas charge transport is improved by the applied field in the whole temperature range. So above 2K, the applied field (\textbf{H}//\textbf{c}) increases $\kappa_{qp}/T$: the observed opposite decrease of $\kappa/T$ in this temperature range has to arise from a strong suppression of $\kappa_{extra}/T$, over-compensating the field increase of $\kappa_{qp}/T$. This points to a sizable extra contribution to heat transport at high temperature, but also to its strong field dependence: hence the need for an extra contribution which cannot be purely originated by phonons. 
The strong decrease of the Lorentz ratio at high temperature under field along the \textbf{c}-axis, as well as the weak sensitivity of this ratio for fields along the (intermediate) \textbf{b}-axis is robust: it has been found on all measured samples. The inset of Fig.\ref{fig:kappa_A12} shows the effect on sample $S^{b}_{150}$ (\textbf{H}//\textbf{c}).

Figure \ref{fig:kappa_A12} shows the thermal conductivity of sample $S^{b}_{150}$ (\textbf{j}$_Q$//\textbf{b}) at 2T, \textbf{H}//\textbf{c} (in the normal state). This sample has a RRR $\approx$10 times larger than sample $S^{c}_{16}$, so a much larger electronic quasiparticle contribution is expected, which should translate in a Lorentz number closer to 1, or even smaller than 1 if inelastic scattering is dominant (as observed in samples $S^{a}_{65}$ and $S^{c}_{110}$). The Lorentz number is indeed much smaller than for sample $S^{c}_{16}$ above 2K (inset of Fig.\ref{fig:kappa_A12}), but not at low temperature: $L/L_{0}$ displays a maximum at around 0.5K, before reaching 1 only for $T\rightarrow 0$. On the raw data, this maximum is well visible at 2T, \textbf{H}//\textbf{c}, whereas the resistivity is monotonously decreasing on cooling. At 6T, the maximum is still visible, although wider and smaller. So for this crystal, which has the largest electronic contribution, $\kappa_{extra}/T$, suppressed also by a field \textbf{H}//\textbf{c}, seems to peak out not only at high temperature, but also in a very low temperature range (where $\kappa_{ph}/T$ is clearly negligible).

\begin{figure}[!ht]
	\centering
		\includegraphics[width=1\textwidth]{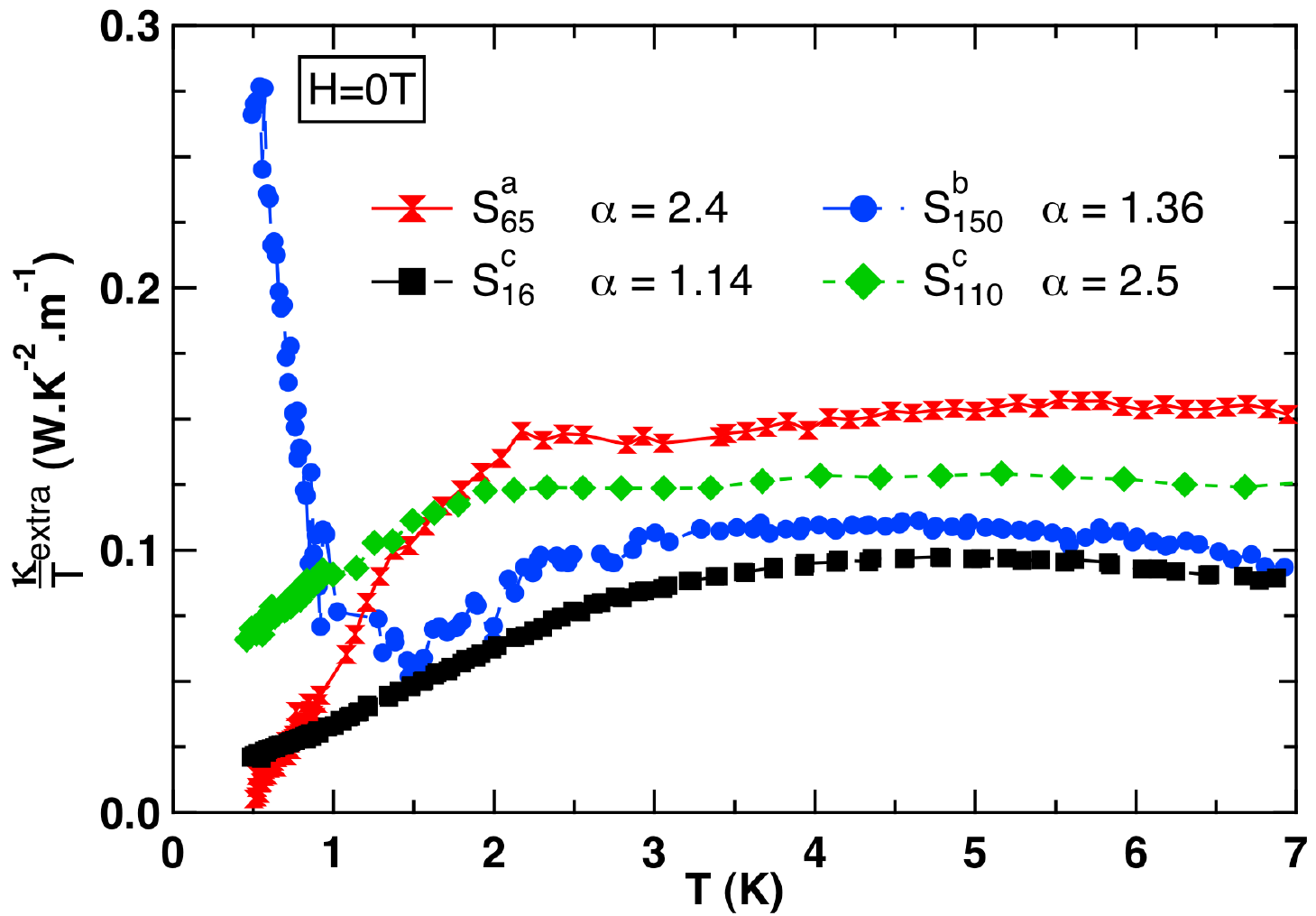}
	\caption{Estimated extra contribution of the thermal conductivity for all samples at zero field. In the paramagnetic state, $\kappa_{extra}/T$ has the same order of magnitude for the four samples whereas it is clearly anisotropic in the ferromagnetic state, and much larger for the heat current along the \textbf{b}-axis.}
	\label{fig:Kautre_0T}
\end{figure}
When the electronic inelastic scattering is negligible compared to other mechanisms, $\kappa_{qp}$ can be simply deduced from the WFL, leading to $\frac{\kappa_{qp}}{T}=\frac{L_0}{\rho}$. A more refined treatment of the electronic quasiparticle contribution is required otherwise. Indeed, inelastic scattering is known to be more efficient for the limitation of thermal conductivity than for the limitation of electrical conductivity, due to the presence of so-called "vertical process" (with energy transfer at small momentum, affecting only the thermal resistivity). A sound estimate of the thermal conductivity is an extended WFL \cite{Wagner1971} of the form 
\begin{eqnarray}
	\frac{\kappa_{qp}}{T}=\frac{L_0}{\rho_0+\alpha(\rho-\rho_0)}
		\label{eq:el_contr_fit}
\end{eqnarray}
$\rho_0$ is the residual resistivity and $\alpha=1+W_{vert}/W_{hor}$, where W$_{vert}$ and W$_{hor}$ are respectively the scattering rates due to inelastic vertical and horizontal processes, assuming that Mathiessen's rule holds. We assumed also that $\alpha$ is sample dependent, larger than 1, but independent of temperature and field. This crude simplification can be valid for temperatures much smaller than the typical energy of the fluctuations responsible for the inelastic scattering, and has already been used for the heavy fermions UPt$_3$ \cite{Lussier1994} or CeRhIn$_5$ \cite{Paglione2005}.

An estimation of the different contributions is given Fig.\ref{fig:kappa_A12} for the sample $S^{b}_{150}$ at 2T with \textbf{H}//\textbf{c}, and the estimated $\kappa_{qp}/T$ is shown for $\alpha=1$ and $\alpha=2$. It appears that the low temperature contribution (below 1K) is little sensitive to the value of $\alpha$, as expected from the small inelastic term of the resistivity below 0.5K, and that larger $\alpha$ values can only enhance the high temperature contribution. Thus, our estimation of $\kappa_{extra}/T$ ($\alpha=1.36$ for this sample), is a lower bound of this contribution.

\begin{figure}[!ht]
	\centering
		\includegraphics[width=1\textwidth]{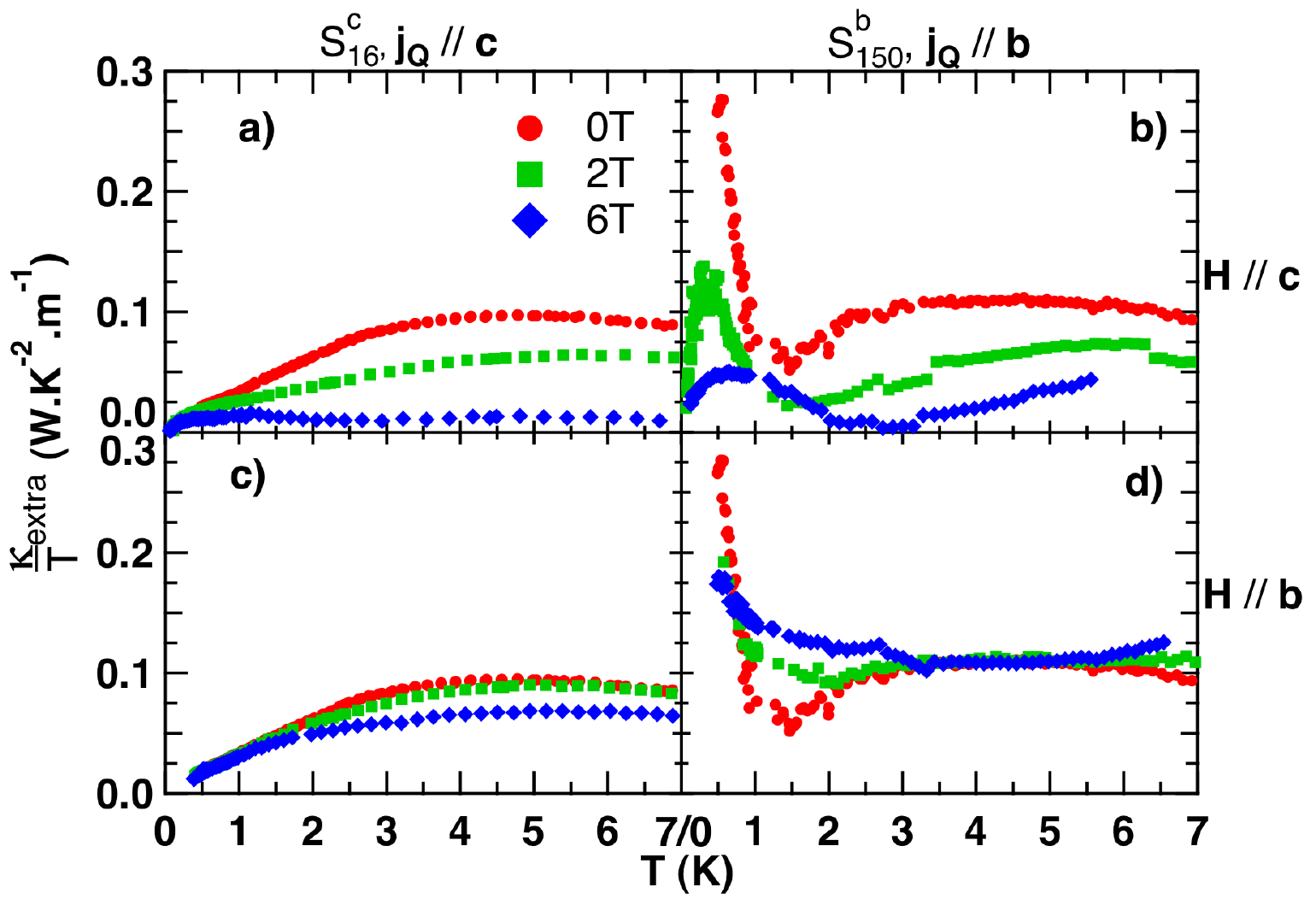}
	\caption{Evolution of $\kappa_{extra}$ of the thermal conductivity for samples $S^{b}_{150}$ (panels \textbf{a)} and \textbf{c)}) and $S^{c}_{16}$ (panels \textbf{b)} and \textbf{d)}) at 0T, 2T and 6T. The panels \textbf{a)} and \textbf{b)} are for \textbf{H}//\textbf{c}, with strong suppression of this contribution, and the panels \textbf{c)} and \textbf{d)} for \textbf{H}//\textbf{b}, showing far less field dependence, and a broadening of the low temperature contribution of sample $S^{b}_{150}$.}
	\label{fig:Kautre}
\end{figure}
The estimated extra contribution to the thermal conductivity (which cannot be attributed to the electronic or phononic heat transport) is estimated in Fig.\ref{fig:Kautre_0T} using eq.\ref{eq:el_contr_fit}, for all samples at zero field, above $T_{sc}$. The precise temperature dependence of $\kappa_{extra}/T$ depends on the approximation used for $\alpha$ (temperature independent), nevertheless, two robust pieces of information can be extracted: i) $\kappa_{extra}/T$ has approximately the same value at zero field in the paramagnetic state for all samples; ii) below 2K, the extra contribution decreases for all samples (including these of high RRR like sample $S^{c}_{110}$) except for sample $S^{b}_{150}$, which shows an increase below 1.5K (see also the curve at 2T in Fig.\ref{fig:kappa_A12}). So quantitatively, the extra contribution is rather isotropic above 2K, but displays a strong anisotropy below this temperature, suggesting a change of the excitation spectrum responsible for this channel of heat conduction below $T_{Curie}$. 

Figure \ref{fig:Kautre} shows $\kappa_{extra}/T$ at various fields (only for samples $S^{c}_{16}$ and $S^{b}_{150}$): $\kappa_{extra}/T$ is strongly suppressed by fields along the \textbf{c}-axis (panels \textbf{a)} and \textbf{b)}), but weakly affected by fields along the \textbf{b}-axis (panels \textbf{c)} and \textbf{d)}). This is strongly reminiscent of the behavior of the longitudinal fluctuations detected by NMR in this system \cite{Ihara2010}. NMR found enhanced ferromagnetic fluctuations below 8K strongly affected by a field along the \textbf{c}-axis. So it is reasonable to identify this extra contribution to heat transport to the longitudinal spin fluctuations detected by NMR. The new information is that these modes can carry heat, so are not just incoherent local moment fluctuations but propagating modes. We also find that his contribution depends little of the crystallographic orientation above 2K, whereas it is much stronger along the \textbf{b}-axis than along the \textbf{c} or \textbf{a}-axis below this temperature: the spectrum of these magnetic fluctuations seems to change when the ferromagnetic order occurs. 

Let us point out that such a picture of long range magnetic fluctuations extending far above the ordering temperature (NMR detects them up to 100K) is typical of the SCR (Self Consistent Renormalized) model of spin fluctuations in weakly ferromagnetic metals, as opposed to the local moment picture where short range order rapidly disappears above $T_{Curie}$ \cite{Moriya1985}. The SCR theory investigates mainly the effects of spin fluctuations on the scattering rate of the conduction electrons \cite{Ueda1975, Jullien1975}, and not their own contribution to an additional channel for heat transport. Experimentally, in ZrZn$_2$ for example \cite{Smith2008}, the resistivity at 50K is still below 10\,~$\mu\Omega$cm (above 40\,~$\mu\Omega$cm at 7K for the best sample of UCoGe), so that $\kappa_{qp}$ is dominant in the whole temperature range and an extra contribution would not be detectable. This is also true for antiferromagnetic systems like CeRhIn$_5$ \cite{Paglione2005}, where resistivity at 7K is still 7 times smaller than that of UCoGe. 

A most striking feature of our results is the strong anisotropy of the extra contribution appearing below $T_{Curie}$. Such a strong anisotropy has been also seen on the thermal expansion measurements \cite{Gasparini2010}: at $T_{Curie}$, there is a large change along the \textbf{b}-axis, a smaller along the \textbf{c}-axis and almost no change along the \textbf{a}-axis, similarly to our measurements. Both effects could be due to the predicted strong change of the Fermi surface below $T_{Curie}$ \cite{Samsel-Czekala2010}, and its expected feedback on the magnetic excitation spectrum in an itinerant ferromagnet.

UCoGe seems to be a rare case among metallic systems, where the spin fluctuation contribution to the thermal conductivity can be directly discriminated. Magnon contributions in the ordered state have been reported in strong (metallic) ferromagnets (see \cite{Yelon1970} for example), or in insulating materials (\cite{Sologubenko2000, Li2005, Jin2003}). But to the best of our knowledge, no "paramagnon" or spin fluctuation contribution has been reported yet: this may come from the tiny long range correlations of local moments  above the ordering temperature, as opposed to the case of itinerant magnets. In UCoGe, the contribution of spin fluctuations could also be favored by the absence of a gap, despite Ising anisotropy: coupling to the Fermi sea may indeed extend the fluctuation spectrum up to zero frequency (see the strong Ising cases of CeRu$_2$Si$_2$ \cite{Knafo2009} or UGe$_2$ \cite{Huxley2003,Raymond2004}), as opposed to insulating materials. Let us note that a direct extra magnetic contribution due to magnons has been recently reported in the heavy-fermion weak antiferromagnet YbRh$_2$Si$_2$ \cite{Pfau2012}. However, in YbRh$_2$Si$_2$, the contribution is still visible at 60mT \cite{Pfau2012}, above the critical field, so that it may be due also to magnetic fluctuations, like in UCoGe, rather than to real magnons.


In conclusion, thermal transport in UCoGe reveals the presence of spin fluctuations, identified to the longitudinal fluctuations seen by NMR, with isotropic propagation above $T_{Curie}$, and mainly \textbf{b}-axis propagation at lower temperature. It confirms an itinerant limit for the magnetism of this compound, despite the strong Ising anisotropy of its magnetic properties, a key feature for the understanding of its superconducting state. The strong anisotropy of the contribution to thermal transport of these fluctuations below $T_{Curie}$ may emerge from the possible dramatic change of the Fermi surface across $T_{Curie}$ \cite{Samsel-Czekala2010}. It seems to be the first time that such a contribution is identified in metallic systems, and urges for theoretical investigations of direct heat transport by spin fluctuations in metallic weak ferro- or antiferromagnets.

\section*{Acknowledgments}
We are pleased to thank S. Raymond, M.E. Zhitomirsky, A.L. Chernyshev, G. Knebel and C. Lacroix for stimulating and fruitful discussions. This work was supported by the French ANR grants SINUS and the ERC grant "NewHeavyFermions".


\end{document}